\documentclass[dvips]{article}
\usepackage{supertabular,lscape,epsfig}
\usepackage{amssymb}
\usepackage{amsmath}
\DeclareSymbolFont{ppa}{OT1}{ppl}{m}{it}
\DeclareMathSymbol{\vv}{\mathalpha}{ppa}{'166}
\begin{document}
\newcommand{\dd}{\,{\rm d}}
\newcommand{\ie}{{\it i.e.},\,}
\newcommand{\etal}{{\it et~al.\ }}
\newcommand{\eg}{{\it e.g.},\,}
\newcommand{\cf}{{\it cf.\ }}
\newcommand{\vs}{{\it vs.\ }}
\newcommand{\zdot}{\makebox[0pt][l]{.}}
\newcommand{\up}[1]{\ifmmode^{\rm #1}\else$^{\rm #1}$\fi}
\newcommand{\dn}[1]{\ifmmode_{\rm #1}\else$_{\rm #1}$\fi}
\newcommand{\upd}{\up{d}}
\newcommand{\uph}{\up{h}}
\newcommand{\upm}{\up{m}}
\newcommand{\ups}{\up{s}}
\newcommand{\arcd}{\ifmmode^{\circ}\else$^{\circ}$\fi}
\newcommand{\arcm}{\ifmmode{'}\else$'$\fi}
\newcommand{\arcs}{\ifmmode{''}\else$''$\fi}
\newcommand{\MS}{{\rm M}\ifmmode_{\odot}\else$_{\odot}$\fi}
\newcommand{\RS}{{\rm R}\ifmmode_{\odot}\else$_{\odot}$\fi}
\newcommand{\LS}{{\rm L}\ifmmode_{\odot}\else$_{\odot}$\fi}
\newcommand{\Abstract}[2]{{\footnotesize\begin{center}ABSTRACT\end{center}
\vspace{1mm}\par#1\par
\noindent
{~}{\it #2}}}
\newcommand{\TabCap}[2]{\begin{center}\parbox[t]{#1}{\begin{center}
  \small {\spaceskip 2pt plus 1pt minus 1pt T a b l e}
  \refstepcounter{table}\thetable \\[2mm]
  \footnotesize #2 \end{center}}\end{center}}
\newcommand{\TableSep}[2]{\begin{table}[p]\vspace{#1}
\TabCap{#2}\end{table}}
\newcommand{\FigCap}[1]{\footnotesize\par\noindent Fig.\  %
  \refstepcounter{figure}\thefigure. #1\par}
\newcommand{\TableFont}{\footnotesize}
\newcommand{\TableFontIt}{\ttit}
\newcommand{\SetTableFont}[1]{\renewcommand{\TableFont}{#1}}
\newcommand{\MakeTable}[4]{\begin{table}[htb]\TabCap{#2}{#3}
  \begin{center} \TableFont \begin{tabular}{#1} #4
  \end{tabular}\end{center}\end{table}}
\newcommand{\MakeTableSep}[4]{\begin{table}[p]\TabCap{#2}{#3}
  \begin{center} \TableFont \begin{tabular}{#1} #4
  \end{tabular}\end{center}\end{table}}
\newenvironment{references}%
{
\footnotesize \frenchspacing
\renewcommand{\thesection}{}
\renewcommand{\in}{{\rm in }}
\renewcommand{\AA}{Astron.\ Astrophys.}
\newcommand{\AAS}{Astron.~Astrophys.~Suppl.~Ser.}
\newcommand{\ApJ}{Astrophys.\ J.}
\newcommand{\ApJS}{Astrophys.\ J.~Suppl.~Ser.}
\newcommand{\ApJL}{Astrophys.\ J.~Letters}
\newcommand{\AJ}{Astron.\ J.}
\newcommand{\IBVS}{IBVS}
\newcommand{\PASP}{P.A.S.P.}
\newcommand{\Acta}{Acta Astron.}
\newcommand{\MNRAS}{MNRAS}
\renewcommand{\and}{{\rm and }}
\section{{\rm REFERENCES}}
\sloppy \hyphenpenalty10000
\begin{list}{}{\leftmargin1cm\listparindent-1cm
\itemindent\listparindent\parsep0pt\itemsep0pt}}%
{\end{list}\vspace{2mm}}
\def\TYLDA{~}
\newlength{\DW}
\settowidth{\DW}{0}
\newcommand{\dw}{\hspace{\DW}}
\newcommand{\refitem}[5]{\item[]{#1} #2%
\def\REFARG{#3}\ifx\REFARG\TYLDA\else, {\it#3}\fi
\def\REFARG{#4}\ifx\REFARG\TYLDA\else, {\bf#4}\fi
\def\REFARG{#5}\ifx\REFARG\TYLDA\else, {#5}\fi.}
\newcommand{\Section}[1]{\section{#1}}
\newcommand{\Subsection}[1]{\subsection{#1}}
\newcommand{\Acknow}[1]{\par\vspace{5mm}{\bf Acknowledgements.} #1}
\pagestyle{myheadings}
\newfont{\bb}{ptmbi8t at 12pt}
\newcommand{\xrule}{\rule{0pt}{2.5ex}}
\newcommand{\xxrule}{\rule[-1.8ex]{0pt}{4.5ex}}
\def\thefootnote{\fnsymbol{footnote}}
\begin{center}
{\Large\bf
Eclipsing binaries in the open cluster NGC 2243 - II. Absolute
properties of NV~CMa \footnote{This paper includes data obtained with the 
6.5-meter Magellan Telescopes located at Las Campanas Observatory, Chile.}}
\vskip1cm
{\bf
J.~~K~a~l~u~z~n~y$^1$,
~~W.~~P~y~c~h$^1$ ,
~~S.~M.~~Rucinski$^2$
and~~I.~B.~~T~h~o~m~p~s~o~n$^3$}
\vskip3mm
{
  $^1$Nicolaus Copernicus Astronomical Center,
     ul. Bartycka 18, 00-716 Warsaw, Poland\\
     e-mail: (jka,pych@camk.edu.pl)\\
  $^2$David Dunlap Observatory, Department of Astronomy and Astrophysics,
University of Toronto, P.O. Box 360, Richmond Hill, ON L4C 4Y6, Canada\\
     e-mail: (rucinski@astro.utoronto.ca)}\\
  $^3$Carnegie Institution of Washington,
     813 Santa Barbara Street, Pasadena, CA 91101, USA\\
     e-mail: (ian@ociw.edu)\\
\end{center}
\Abstract
{We present echelle spectroscopic data for five eclipsing binary stars
and two giant stars in the field of the open cluster NGC~2243. The
average cluster velocity is determined to be $+60.4\pm 0.6~km~s^{-1}$.
Four of the eclipsing binaries are very likely members of the cluster
based on their observed radial velocities. The absolute parameters of
cluster member NV~CMa are determined by analysing photometric and radial
velocity data. We obtain $1.089\pm 0.010 M_{\odot}$ and $1.221\pm
0.031 R_{\odot}$ for the primary, and $1.069\pm 0.010 M_{\odot}$ and
$1.178\pm 0.037 R_{\odot}$ for the secondary. Both components of the
binary are located on the Main Sequence, about 1~mag. below the
turn-off point on the cluster color-magnitude diagram. Using model
age-luminosity and age-radius relations we obtain $4.35\pm0.25$~Gyr
for the age of NV~CMa. The derived age is, however, very sensitive
to the adopted metallicity of the cluster. We demonstrate that a
meaningful determination of the ages of objects like NV~CMa based on
evolutionary models is possible only if their metallicity is know with
a relative accuracy of a few percent. The distance moduli calculated
for the components of NV~CMa agree closely with each other, and imply
an apparent distance modulus of the cluster of $(m-M)_{\rm V}=13.25\pm
0.08$.
}
{Stars: binaries: eclipsing, binaries -- stars: individual: NV CMa -- open
clusters and associations: individual: NGC~2243}
\Section{Introduction}
The field of the intermediate-age open cluster NGC~2243 contains 5
known detached eclipsing binaries. Four of these are likely cluster
members based on their photometric properties (Kaluzny et~al. 2006;
hereafter KKTS). The components of these systems should have the same
age, metallicity and heliocentric distance and the determination of
their absolute parameters can provide an interesting test of
evolutionary models of low mass stars. Moreover, through the use of the
surface brightness method, one may obtain a direct measure of the
cluster distance.

This paper is focused on the determination of the absolute properties
of NV~CMa, an eclipsing binary located in the central area of NGC~2243.
In addition we use spectroscopic observations to check on
the membership status of 4 other binaries in the cluster field.
\Section{Spectroscopic Observations and Reductions}
Spectroscopic observations of NGC~2243 stars were carried out
with the MIKE echelle spectrograph (Bernstein et~al. 2003) on the Magellan~II
(Clay) telescope of the Las Campanas Observatory.
The data were collected
during observing runs in 2004 October and 2005 September.
For this analysis we use data 
obtained with the blue channel of MIKE
covering the range from 400 to 500 nm
with a resolving power of $\lambda / \Delta \lambda \approx 38,000$.
All of the observations were obtained with a $0.7\times 5.0$ arcsec slit
and with $2\times 2$ pixel binning. At 4380 \AA\ the resolution was
$\sim$2.7 pixels at a scale of 0.043~\AA/pixel. 
The seeing ranged from 0.7 to 1.1~arcsec.
%at 4380A 1pix=0.043A FWHM=2.7 -> R=4380/(2.7*0.043=37758)
The spectra were first processed using a pipeline developed by Dan
Kelson following the formalism of Kelson (2003, 2006) and then analysed
further using standard tasks in the IRAF/Echelle package\footnote{IRAF
is distributed by the National Optical Astronomy Observatories, which
are operated by the Association of Universities for Research in
Astronomy, Inc., under cooperative agreement with the NSF.}. Each of
the final individual spectra consisted of two 600~s exposures
interlaced with an exposure of a thorium-argon lamp. We obtained 12
spectra of NV~CMa. The average signal-to-noise ratios range from 25
at shorter wavelengths to 50 at  longer wavelengths.
%   The wavelength range is given few sentences earlier   %
In addition to observations of NV~CMa we also
obtained single spectra of the eclipsing binaries V4=NS~CMa,
V5=NX~CMa, V7 and V9 (names of variables as in KKTS), and for the red
giants H3110 and H4115 (Hawarden 1975) located in the central part of
the cluster field.
\subsection{Spectroscopic Orbit of NV~CMa}
Radial velocities of the components of NV~CMa were measured by
cross-correlation with the FXCOR task in IRAF, using observations of
HD~33256 as a template. According to Nordstrom et~al. (2004), HD~33256
has $V_{rad}=10.1\pm 0.2~km~s^{-1}$ and a projected rotational velocity
$V \sin i=10~km~s^{-1}$. With $B-V=0.45$ and $[{\rm Fe/H}]=-0.5$, it
provides a good match for the color index and metallicity of the
binary. The template was observed with the same instrumental
configuration as the variable. The correlation peaks were measured
to have a FWHM of about 70 $~km~s^{-1}$. All spectra showed the peaks to be
separated by more than 1.4 FWHM, and the peaks were measured
simultaneously with the FXCOR package. The correlation was measured only
from the metal lines, excluding the H$\beta$,
H$\gamma$, and H$\delta$ hydrogen lines. Our velocity
measurements for NV~CMa are listed in Table~1 and are shown in Fig.~1.
A Keplerian orbit was fitted to the observations by fixing the period
and epoch based on the precise ephemeris established by KKTS:
\begin{equation}
Min I = HJD~244 8663.70748(5) + 1.18851590(2)
\end{equation}
We assumed a circular orbit based on the photometric data.
This assumption is further supported by the relatively short orbital period
of the variable compared to the cluster age
of about 3.8~Gyr (Anthony-Twarog, Atwell \& Twarog 2005).
For an age of 3.8~Gyr all NGC~2243 binaries with periods shorter than
a few days are expected to have circularized orbits (Mathieu 2005).
The adjustable parameters in the orbital solution were the velocity
semi-amplitudes ($K_{1}$ and $K_{2}$) and the systemic velocity
($\gamma$). The fit was performed using the GAUSSFIT task within 
IRAF/STSDAS.
The derived parameters of the spectroscopic orbit are listed in
Table~2\footnote{Through this paper we adopt the following values of
constants: $R_{\odot}=6.95508E5$~km,
$M_{\odot}=1.9891E30$~kg, $G=6.67259E-11$ ${\rm m^{3}kg^{-1}s^{-1}}$.}.
The systemic velocity of NV~CMa agrees within the measurement errors with
the radial velocity of the cluster as estimated in the next subsection.
\begin{table}
\centering
\caption{\small Radial velocities of NV~CMa and residuals from the adopted
spectroscopic orbit}
{\small
\begin{tabular}{llrrrrrr}
\hline
Phase&HJD-2453000& $RV_{1}$& $\sigma_{\rm RV1}$& $RV_{2}$ & $\sigma_{\rm RV2}$&${\rm (O-C)_{A}}$& ${\rm (O-C)_{B}}$\\
\hline
 0.2634& 724.7208& -67.20  &2.77& 192.23&  2.69& -0.81&  0.12\\
 0.3178& 636.8353& -56.61  &1.50& 182.27&  1.51& -1.25&  1.40\\
 0.3375& 636.8587& -45.65  &1.51& 173.65&  2.04&  2.26&  0.36\\
 0.3557& 636.8803& -40.76  &1.31& 161.51&  1.80& -1.23& -3.24\\
 0.4344& 282.7961&  12.29  &1.26& 114.13&  1.18&  2.09&  0.00\\
 0.5930& 281.7961& 134.08  &1.06& -11.81&  1.10&  1.47& -1.32\\
 0.6100& 281.8163& 143.47  &1.02& -22.17&  1.10& -0.17& -0.45\\
 0.6259& 281.8353& 152.98  &1.18& -31.36&  1.14& -0.13&  0.00\\
 0.6422& 281.8546& 160.28  &1.22& -40.58&  1.26& -1.59& -0.30\\
 0.6582& 281.8736& 170.66  &1.31& -49.45&  1.39&  1.21& -1.45\\
 0.8310& 633.8797& 173.11  &1.26& -51.55&  1.55& -0.85&  1.03\\
 0.8401& 633.8905& 170.83  &1.35& -46.98&  1.47&  0.63&  1.77\\
\hline
\end{tabular}}
\end{table}
\begin{table}
\centering
\caption{\small Orbital parameters of NV~CMa.}
{\small
\begin{tabular}{llrrrrrr}
\hline
Parameter & Value \\
\hline
$P$ (days) & 1.18851590(fixed) \\
$T_{0}$~(HJD-244 0000) & 8663.70748(fixed) \\
$\gamma~(km~s^{-1})$ & 61.70$\pm$ 0.30 \\
$e$    & 0(fixed) \\
$K_{1}~(km~s^{-1})$  & 128.55$\pm$0.52 \\
$K_{2}~(km~s^{-1})$  & 130.87$\pm$0.55 \\
Derived quantities: & \\
$A~sin~ i$~($R_{\odot}$) & 6.096$\pm$0.018\\
$M_{1}~sin^{3}~i$~($M_{\odot}$) & 1.084$\pm$0.010 \\
$M_{2}~sin^{3}~i~$($M_{\odot}$) & 1.065$\pm$0.010 \\
Other quantities: & \\
$\sigma_{1}~(km~s^{-1})$ & 1.37 \\
$\sigma_{2}~(km~s^{-1})$ & 1.37 \\
\hline
\end{tabular}}
\end{table}
\subsection{Analysis of Broadening Functions}
We have analysed the spectra of NV~CMa using code based on
the broadening function (BF) formalism (Rucinski 2002).
The BF analysis lets us study the effects of the spectral 
line broadening and orbital splitting even for relatively
complicated profiles of spectral lines. This macroscopic
velocity information is obtained 
regardless of the parameters of the photosphere such as the
temperature, pressure etc. It retains strict linearity
in reproduction of the individual contributions to the
broadening profile, i.e.\ the individual luminosities of
components can be simply estimated from the strengths of
the respective peaks in the broadening profile.

In the case of detached binary systems the dominating
effect on the observed broadening of the spectral lines
comes from the rotation of the components induced by the orbital motion.
If we make the approximation of rigid rotation of perfectly spherical
stars we can integrate the light over the visible hemispheres analytically.
The resulting theoretical rotational BF has a shape described
by:
\begin{eqnarray}
BF_{rot}(v) & = & A [(1-\beta)\sqrt{1-a^2} + {\frac{\pi}{4}} \beta (1 - a^2)] + C
\\ \nonumber \\
a & = & {\frac{v-v_{rad}}{v_{rot}\sin{i}}} \nonumber
\end{eqnarray}
where $A$ is a normalization constant, $\beta$ is the linear limb darkening
coefficient, $C$ is the continuum level, $v_{rad}$ is the radial velocity of
the center of mass of the star, $v_{rot}$ is the linear velocity on the
equator of the rotating star, and $i$ is the inclination angle between
the axis of rotation and the direction to the observer.

We have extracted BFs from all of the spectra of NV~CMa and have
conducted nonlinear least squares fitting of the model profile to the
observed BFs to measure the radial and projected rotational velocities
of the components of NV~CMa. In both steps of the calculation we have
used our programs based on procedures from the GNU Scientific Library.

As it will be shown in Section 3, the components of NV~CMa are almost
spherical and there are no signatures of spot activity. Our model
profile is then the sum of two theoretical rotational BFs convolved
with a Gaussian with a standard deviation of $15~km~s^{-1}$. The
Gaussian represents effects of the 
instrumental resolution and is a part of the BF method in which modest
smoothing (adjusted to the width of the spectrograph slit image)
is applied to the BFs. We used the spectra in
the wavelength range from $400~nm$ to $495~nm$. This range roughly
corresponds to the B passband in the UBV photometric system. Since the
shape of the rotational BF only weakly depends on the limb darkening
coefficient, we have adopted a constant $\beta = 0.63$ as derived from
the $(B-V,~\beta)$ relation of van Hamme (1993). The four parameters
$A$, $C$, $v_{rad}$, and $v_{rot}\sin{i}$ were simultaneously adjusted
in the fitting procedure. Figure~2 presents an example of fitting
the model to the BF calculated for a spectrum taken at orbital phase
0.61. A linear least squares fit to the measured radial velocities
gives the following orbital solution: $\gamma= 61.47 \pm
0.14~km~s^{-1}$, $K_1= 130.93 \pm 0.59~km~s^{-1}$, $K_2= 128.62 \pm
0.42~km~s^{-1}$, which lies well within formal errors from the result
obtained using the cross-correlation analysis. The RMS error of the fit is
$0.93~km~s^{-1}$ for both components. We measure the projected
rotational velocities to be $V_1\sin{i}= 51.7 \pm 1.4~km~s^{-1}$ and
$V_2\sin{i}= 52.4 \pm 2.5~km~s^{-1}$. Note that these rotational
velocities are very close to those expected if the components are in
synchronous rotation in a circular orbit, adopting the period from
Table 2 and the stellar radii from Table 1 ($V_1\sin{i}= 52.0 \pm
1.3~km~s^{-1}$ and $V_2\sin{i}= 50.1 \pm 1.6~km~s^{-1}$).
[SMR: Strictly speaking, we should subtract quadratically the 
Gaussian smoothing of sigma = 15 km/s. Then one gets 49.5 and
50.2 km/s, in almost perfect agreement with the expectations.]  

The integral of a BF profile is proportional to the total flux from a star.
Since the absolute measurement of this flux requires a perfect match between
the template and object spectra we did not attempt to conduct such an analysis.
However, given that the effective temperatures of the components of
NV~CMa are very similar, the integrals of the BF profiles are proportional to
the monochromatic luminosities of the two components.
An analytical integration of the profiles
leads to the conclusion that the ratio of the luminosities of stars
with identical spectral types is proportional to the ratio
of the products $A \cdot v_{rot}\sin{i}$ for each of the system
components. From a total of 11 spectra with well defined BF profiles we
obtained $L_{1B}/L_{2B}=1.087\pm 0.011$ ($rms=0.034$), in good agreement
with the values measured from the photometric observations (see Section 3).
\subsection{Velocities of other objects}
Table~2 lists the radial velocities derived for two red giants and four
other eclipsing binaries from the cluster field. The velocities of
H3110 and H4115 are close to the velocities of two other cluster red
giants observed by Gratton (1982) who measured $V_{rad}=+62~km~s^{-1}$
and $V_{rad}=+60~km~s^{-1}$ for H4110 and H4209, respectively. Based on
these four stars we estimate the radial velocity of the cluster to be
$V_{rad}=+60.4\pm 0.6~km~s^{-1}$.

Two velocity peaks are seen in the cross-correlation functions of V4,
V5, and V7. The mean velocity for the two peaks for V4 is
$210.7~km~s^{-1}$, well above the cluster velocity. We conclude that V4
is a field binary star not related to the cluster. In the case of V5, two
peaks of very similar shape and height are seen in the
cross-correlation function. The average velocity of the two components
is $59.1~km~s^{-1}$. V5 is a very likely member of the cluster with a
mass ratio close to to unity. Three peaks show up in the
cross-correlation function obtained for V7, one at $22.8~km~s^{-1}$,
one at $96.5~km~s^{-1}$ and a wing to that component at
$136.8~km~s^{-1}$. The mean of first two is $59.6~km~s^{-1}$, very
similar to the cluster average velocity. It is possible that the
component with highest velocity is caused by contaminating light from
the close visual companion of V7 whose light leaked through the slit
during observations (see finding chart in KKTS). For V9 only one
strong peak was detected in the cross-correlation function at a velocity
of $65.4~km~s^{-1}$, consistent with cluster membership. It is possible
that this binary was observed close to conjunction, and further
observations are needed to see if the secondary component can be
detected in the cross-correlation function. In conclusion, variables
V5, V7 and V9 are likely members of NGC~2243. These three binaries are
promising targets for detailed observations aimed at the determination of
the parameters of their components.
\begin{table}
\centering
\caption{\small Radial velocities of stars from the field of NGC~2243 }
{\small
\begin{tabular}{lllll}
\hline
Name & HJD-2450000& ${\rm RV_{1}}$ & ${\rm RV_{2}}$ & member?\\
\hline
H3110 & 3281.9127  & 59.4(1.0)  &             & yes\\
H4115 & 3281.9285  & 60.1(1.1)  &             & yes \\
V4    & 3635.9006  & 117.5(2.4) &285.9(9.6)   & no\\
V5    & 3638.8430  & 23.50(0.58) & 94.61(0.56)& yes\\
V7    & 3638.8217  & 96.88(0.80) & 22.8(5.3)  & yes\\
%V7=old lc8
V9    & 3638.8711  & 65.45(0.68) &            & yes\\
%V9=old lc17
\hline
\end{tabular}}
\end{table}
\Section{Light Curve Solution of NV~CMa}
We have analysed the $BV$ light curves of NV~CMa obtained by KKTS using
the Wilson-Deviney model (Wilson \& Deviney 1971) as implemented in the
light-curve analysis program MINGA\footnote{ MINGA is available at
http://ftp.camk.edu.pl/camk/tomek/Minga/} (Plewa 1988). The mass-ratio
of the binary was fixed at the spectroscopic value of $m_2/m_1 = 0.9825
\pm 0.0018$. The gravity darkening exponents and bolometric albedos
were fixed at 0.32 and 0.5, respectively. The linear darkening
coefficients were adopted for the $B$ and $V$ filters from van Hamme
(1993) for an assumed metallicity of ${\rm [Fe/H]=-0.49}$ (Gratton \&
Contarini 1994; Friel et~al. 2002). An interpolation routine in the
PHOEBE package (Pr\^sa \& Zwitter 2005) was used to get values
corresponding to the adopted effective temperatures of the two
components. The light curves used in the analysis are shown in
Fig.~3. They contain 121 points in the $B$ band and 446 points in the
$V$ band. Outside of the eclipses, we used normal points formed by
averaging 3 to 7 individual observations. There is no evidence for
totality in any of the eclipses. The primary and secondary eclipses
have very similar depths, implying similar surface
brightness and in turn similar effective temperatures for the two
components. The light curve is symmetric, indicating that
its shape is not noticeably affected by spot activity.

The average color index near quadrature is $<B-V>_{max}=0.439\pm
0.020$. The quoted uncertainty includes an external error arising from
the photometric calibration. Adopting $E(B-V)=0.055\pm 0.004$
(Anthony-Twarog et~al. 2005) we obtain an unreddened color index at
maximum light of $(B-V)_{0}=0.384$. Using the empirical
calibration of Ram{\'i}rez \& Mel{\'e}ndez (2005) we get an effective
temperature of the primary component of NV~CMa of $T_{1}=6522\pm 129$~K.
The uncertainty includes the formal uncertainty in the temperature
calibration as well as the uncertainty of the color index. As it is
shown below, the effective temperatures of the components do not differ
by more than about 30~K and the color index at maximum light can be
safely adopted as the color index of the primary component.

The following parameters were adjustable in the light curve solution:
the orbital inclination $i$, the non-dimensional potentials $\Omega_{1}$
and $\Omega_{2}$, the effective temperature of the secondary $T_{2}$,
and the relative luminosity of the primary $L_{1}(V;B)$. For a fixed
value of the mass ratio $q$ the potentials $\Omega_{1}$ and
$\Omega_{2}$ directly determine the relative radii of the components
$r_{1}$ and $r_{2}$. In the following discussion we list "equal volume"
mean radii of the components. Our finally adopted solution (see below)
implies that for both components of NV~CMa the difference between
"polar" and "point" radii amounts to about 2\% \footnote{The "polar
radii" is a radius toward the stellar pole while "point radii" is the
radius toward the Lagrangian point L1 of the binary orbit.}. An
unconstrained light curve solution obtained with MINGA is listed in
Table~4. It is worth noting at this point that the errors returned by
MINGA take into account correlations between all fitted parameters.
One should note the rather large uncertainties of the derived relative
radii and luminosities. This is not an unexpected result given the
partial eclipses (Irwin 1962). An additional complication is
that the effective temperatures of the components of NV~CMa are very
similar.

The accuracy of the fit can be significantly improved by using information
about the light ratio of the two components derived from the
spectroscopic data. A grid of solutions was calculated with a fixed
value of $\Omega_{1}$ (this is equivalent to fixing the value of the
radius $<r_{1}>$). The result is presented graphically in Fig.~3 which
shows the calculated values of $L_{1B}/L_{2B}$ and $<r_{2}>$ as a
function of the assumed value of $<r_{1}>$. There is a strong
anti-correlation between the calculated value of $<r_{2}>$ and the
assumed value of $<r_{1}>$.
The solutions fulfill the condition $L_{1B}/L_{2B}=1.087\pm 0.011$
only for a very narrow range of radii, from $<r_{1}>=0.2001$ to
$<r_{1}>=0.2014$. However, the uncertainty in $<r_{1}>$ obtained this
way is severely underestimated. To get a realistic estimate of the
errors in the fitted parameters one has to take into account correlations
between them. To attain that goal we derived solutions for 3 fixed
values of $L_{1B}/L_{2B}$ spanning the range $1.087\pm 0.011$. The
adjustable parameters were $i$, $\Omega_{1}$, $\Omega_{2}$, $T_{2}$ and
$(L_{1V}/L_{2V})$. The final light curve solution along with the
uncertainties of all fitted parameters is listed in Table~5. Figure~5 shows
the residuals corresponding to this solution.
%{\bf what is the value of $T_{1}$ for this solution?}
%
%These solutions imply also $<r_{2}=$0.1938-1927 and $T_{2}$=5606
%
%In addition to demonstrate how poorly constrained is we
%plotted in this figure also dependence of reduced $\chi^{2}$
%on $<r_{1}>$.
%
%
\begin{table}
\centering
\caption{\small An unconstrained light curve solution}
{\small
\begin{tabular}{ll}
\hline
Parameter & Value \\
\hline
$i$~(deg) & 87.08 $\pm$ 0.95 \\
$\Omega_{1}$ & 5.959 $\pm$ 0.332 \\
$\Omega_{2}$ & 6.141 $\pm$ 0.392 \\
$T_{1}$~(K) & 6522 (fixed) \\
$T_{2}$~(K) & 6512 $\pm$ 123 \\
$(L_{1V}/L_{2V})$ & 1.110 $\pm$ 0.017 \\
$(L_{1B}/L_{2B})$ & 1.112 $\pm$ 0.024 \\
$<r_{1}>$ & 0.203 $\pm$ 0.013\\
$<r_{1}>$ & 0.193 $\pm$ 0.014\\
rms~(V)~(mag) & 0.008 \\
rms~(B)~(mag) & 0.006 \\
\hline
\end{tabular}}
\end{table}
\begin{table}
\centering
\caption{\small Constrained light curve solution with
$(L_{1B}/L_{2B})=1.087 \pm 0.011$}.
{\small
\begin{tabular}{ll}
\hline
Parameter & Value \\
\hline
$i$~(deg) & 87.09 $\pm$ 0.66 \\
$\Omega_{1}$ & 6.010 $\pm$ 0.136 \\
$\Omega_{2}$ & 6.128 $\pm$ 0.155 \\
$T_{1}$~(K) & 6522 (fixed) \\
$T_{2}$~(K) & 6506 $\pm$ 75 \\
$(L_{1V}/L_{2V})$ & 1.088 $\pm$ 0.024 \\
$<r_{1}>$ & 0.200 $\pm$ 0.005\\
$<r_{2}>$ & 0.193 $\pm$ 0.006\\
rms~(B)~(mag) & 0.008 \\
rms~(V)~(mag) & 0.006 \\
\hline
\end{tabular}}
\end{table}
\section{Absolute properties}
The absolute parameters of NV~CMa obtained from our spectroscopic and
photometric analysis are given in Table~6. The errors in the
temperatures include all sources of uncertainties. The absolute visual
magnitudes $M_{\rm V}$ were calculated using bolometric corrections
derived from relations presented by VandenBerg \& Clem (2003). The
observed visual magnitudes derived from the light curve solutions are
$V{_1}=17.107\pm 0.023$ and $V{_2}=17.199\pm 0.024$, where the
uncertainties include the errors of the photometric zero point. For
the $B$ band we obtain $B{_1}=17.549\pm 0.021$ and $B{_2}=17.640\pm
0.021$. Figure~6 shows the location of the individual components of
NV~CMa on a color-magnitude diagram at the turnoff region of NGC~2243.
Note that NV~CMa lies on the binary sequence in this cluster
(Bonifazi et al. 1990) while the individual components lie on the
sequence of single stars, further confirmation of the cluster
membership of NV~CMa. Using these values for the observed and absolute
visual magnitudes one obtains an apparent distance modulus $(m-M)_{\rm
V}=13.25\pm0.10$ and $(m-M)_{\rm V}=13.25\pm0.15$ for the primary and
the secondary components of the binary, respectively, with an average
value of $(m-M)_{\rm V}=13.25\pm0.08$. This value is in good agreement
with recent determinations of $(m-M)_{\rm V}=13.15\pm0.1$ obtained from
the isochrone fitting method by both Anthony-Twarog et~al. (2005; they
adopted ${\rm [Fe/H]}=-0.57$) and VandenBerg et~al. (2006; they adopted
${\rm [Fe/H]}=-0.61$).

\subsection{The age}
It is possible to estimate ages of the components of the binary by
using theoretical age-luminosity relations. It is worth noting that
age-luminosity relations based on stellar models are unaffected by
uncertainties associated with model isochrones relating $T_{eff}$ and
$L_{bol}$ to color index and absolute magnitude in a selected band. In
particular, they are unaffected by the way in which models treat
sub-photospheric convection in low mass stars. In Fig.~7 we show age
versus luminosity relations based on evolutionary tracks recently
published by VandenBerg et~al. (2006). The tracks were derived from a
set of isochrones calculated with the program $vriso$ distributed
with the model grids.
Models for ${\rm [Fe/H]}=-0.525$ (see the following section) and ${\rm
[\alpha/Fe]}=0.3$ were used. Horizontal lines in Fig.~7 mark $L\pm
\sigma_{L}$ ranges for a given component. The intersections of
age-luminosity relations with lines marking $1~\sigma$ limits of the
luminosities give limits on the age for a given mass. The $1~\sigma$ age
limits are $4.00<t_{1}<5.8$~Gyr and $4.1<t_{2}<6.3$~Gyr for the primary
and secondary components of NV~CMa, respectively. The large ranges
result mainly from the fact that both components are still in a
relatively slow phase of their evolution: as can be seen in Fig.~6 they
are located about 1~$mag.$ below the turn-off point on the cluster
color-magnitude diagram. One may also note from Fig.~7 that the errors
of the luminosities and masses contribute an equal 0.6~Gyr to the total
uncertainty in the estimated ages.

Figure~8 shows the time dependence of the radius, also a sensitive
diagnostic for the age, using the same set of evolutionary models. As
before, the solid lines correspond to evolutionary tracks for the
masses of components of NV~CMa. The measured radii are indicated by the
horizontal lines spanning the range $\pm 1~\sigma$.

>From Fig.~8 we derive $1~\sigma$ limits on the ages of the components
of NV~CMa of $3.2<t_{1} < 4.6$~Gyr and $3.2<t_{1}<4.9$~Gyr. These
limits are consistent with these derived from the age-luminosity
relations. The age-radius relations suggest slightly lower ages in
comparison with the age-luminosity relation for both components.
The overlap of the age estimates implies an age for NV~CMa of
approximately 4.1 to 4.6 Gyr.

\subsection{Metallicity}
The age estimates presented in the previous section suffer from a 
potential systematic error arising from the adopted 
metallicity of NV~CMa.
There are three modern determinations of the cluster metallicity.
Gratton \& Contarrini (1994) obtained high resolution spectra of two
cluster giants and derived ${\rm [Fe/H]}=-0.48\pm 0.15$.
Friel et~al. (2002) used medium resolution spectra of 9 stars to derive
${\rm [Fe/H]}=-0.49\pm 0.05$. Finally, Anthony-Twarog et~al. (2005)
employed $uvbyCaH\beta$ photometry to obtain ${\rm [Fe/H]}=-0.57\pm 0.03$.
The weighted average of these three determinations
gives ${\rm [Fe/H]}=-0.547\pm 0.025$. This value is very close to
${\rm [Fe/H]=-0.525}$ for which we extracted the model relations used
above. However, the estimated age 
of NV CMa is very sensitive to the adopted metallicity. For example,
using models for ${\rm [Fe/H]}=-0.397$ and ${\rm [Fe/H]}=-0.606$ we
obtain ages $t=5.425\pm 0.025$~Gyr and $t=3.85\pm 0.45$~Gyr,
respectively, where the very small formal errors just indicate the
marginal overlap between the age ranges obtained from age-luminosity
and age-radius relations. In summary, for the masses and ages relevant
to the present discussion, the metallicity of the analysed stars has to
be known with a relative accuracy of a few percent to allow meaningful
comparison of observational data with the models. This is illustrated
in Fig.~9 which shows age-luminosity and age-radius relations for a
star with $m=1.089 m_{\odot}$ and for 3 values of metallicity. The
plotted relations are based on models by VandenBerg et~al. (2006) for
${\rm [\alpha /Fe]}=+0.3$. At a given age the separation between
relations for close, but different, metallicities exceeds the
uncertainties of the luminosities and radii of components of NV~CMa
obtained in our analysis presented above.
\begin{table}
\centering
\caption{\small Absolute parameters of NV~CMa}
{\small
\begin{tabular}{llrrrrrr}
\hline
Parameter & Value \\
\hline
$A$~($R_{\odot}$)               & 6.104$\pm$0.018\\
$M_{1}$~($M_{\odot}$)  & 1.089$\pm$0.010 \\
$M_{2}$~($M_{\odot}$)  & 1.069$\pm$0.010 \\
$R_{1}$~($R_{\odot}$)  & 1.221$\pm$0.031 \\
$R_{2}$~($R_{\odot}$)  & 1.178$\pm$0.037 \\
$T_{1}$~(K)            & 6522$\pm$129 \\
$T_{2}$~(K)            & 6506$\pm$149 \\
$Lbol_{1}$($L_{\odot}$)    & 2.42$\pm$0.23\\
$Lbol_{2}$($L_{\odot}$     & 2.23$\pm$0.25\\
$M_{\rm V1}~(mag)$          & 3.86$\pm$0.10\\
$M_{\rm V2}~(mag)$          & 3.95$\pm$0.15\\
\hline
\end{tabular}}
\end{table}
\Section{Discussion and summary}
The analysis of photometric and spectroscopic observations of the
eclipsing binary NV~CMA has allowed us to determine the orbital
parameters and physical properties of the component stars.
Our determinations have formal uncertainties of 1\% in the masses
and 3\% in radii. Uncomfortably large uncertainties in the radii result
from the degeneracy of the light curve solution for this partially
eclipsing system. Comparison with model tracks for $\rm{[Fe/H]}=-0.525$
give an age of 4.1 to 4.6 Gyr for the binary and consequently
for the cluster. This determination suffers
from a substantial systematic error
related to the uncertain metallicity of the cluster.
For the relevant range of
stellar masses and ages an uncertainty in the metallicity of 0.1~dex
leads to an uncertainty in the estimated age of about 0.8~Gyr.

There is still room for obtaining a better determination of the
parameters of NV~CMa, including its age. First of all,
it is straightforward to derive masses of the components
with an accuracy of 0.5\% or even better. We used 12 spectra only
and additional radial velocity data obtained near quadrature will lead
to an improvement in the estimates of the masses of the components.
It is also possible to obtain
better light curves than these used in the present analysis;
an improvement is possible in both the quality and the phase coverage.
In particular, our photometry for the $B$ band had
poor coverage inside the primary eclipse. 
Improved light curves would better constrain the parameters of the
components obtained in the light curve solution.
And last, but not least, the age determination of NV~CMa and of other
cluster binaries would benefit enormously from accurate metallicity
determinations, preferably
based on high resolution spectroscopy of an extended 
sample of cluster member stars.

The analysis of single spectra obtained for four other eclipsing
binaries in the cluster field indicates that three of them are radial
velocity members of NGC~2243. The fourth star is definitely a
non-member. Further observations of member binaries would allow a
better constraint on the cluster age as well as a test of evolutionary
models of low-mass stars with $\rm{[Fe/H]}\approx -0.5$.
%
%fig 1
\begin{figure}[htb]
\centerline{\includegraphics[height=80mm,width=120mm]{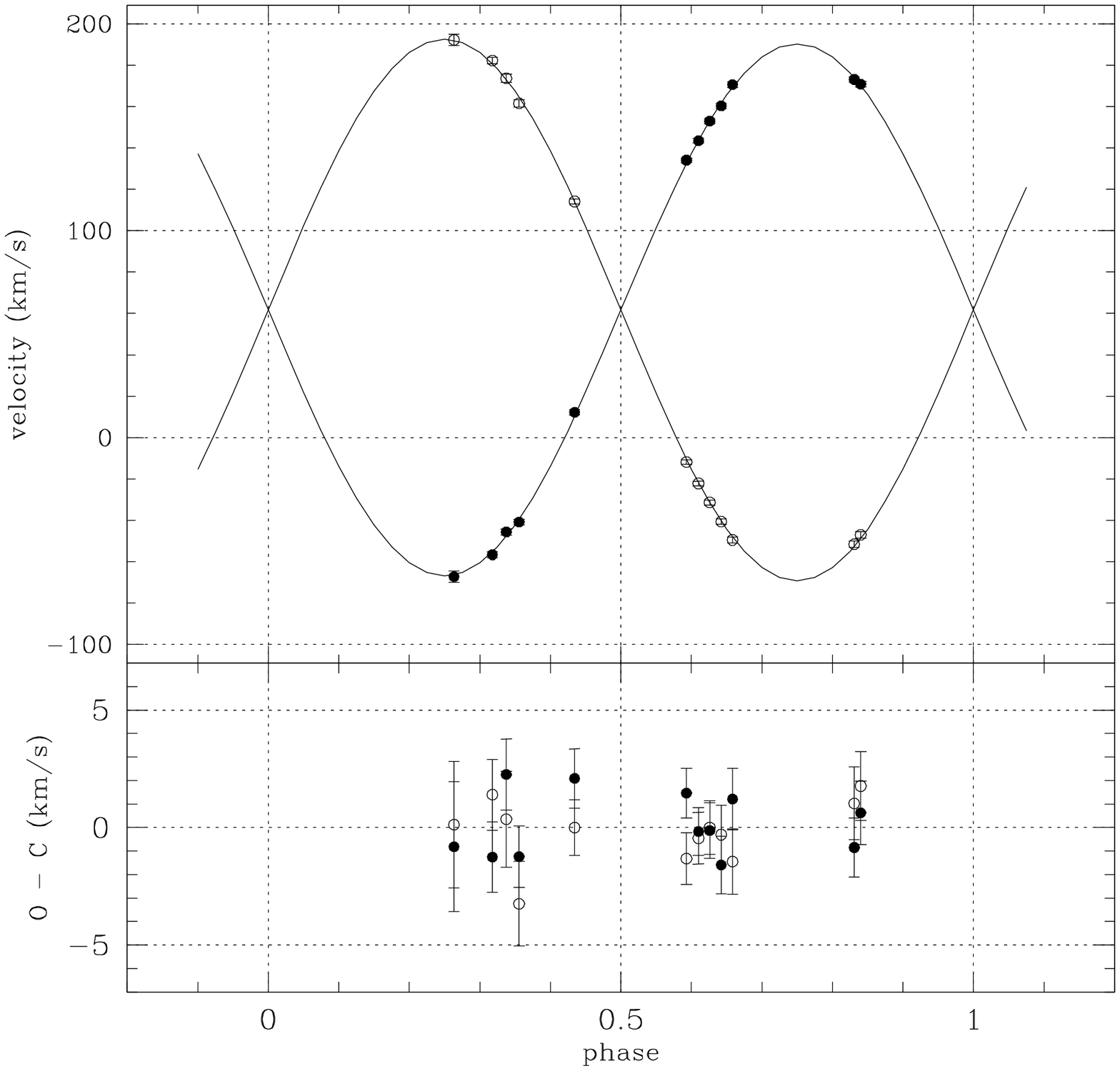}}
\FigCap{Spectroscopic observations and adopted orbit for NV~CMa.}
\end{figure}
%
%fig 2
\begin{figure}[htb]
\centerline{\includegraphics[height=90mm,width=120mm]{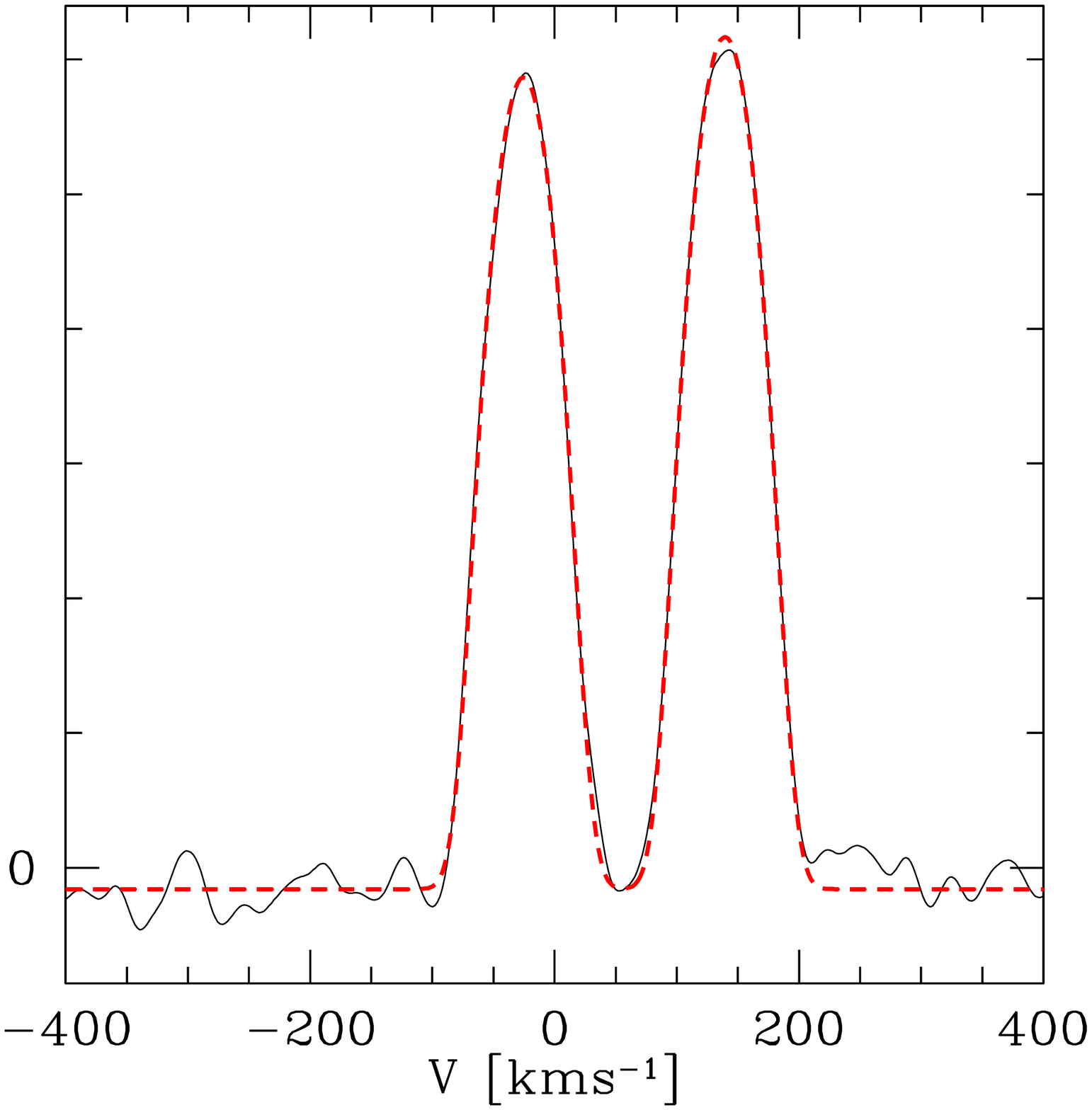}}
\FigCap{Broadening Function extracted from a spectrum of NV~CMa obtained
at orbital phase 0.61 (solid line). The dashed line shows the fit of a
model BF to the observed one.}
\end{figure}
%
%fig 3
\begin{figure}[htb]
\centerline{\includegraphics[height=80mm,width=120mm]{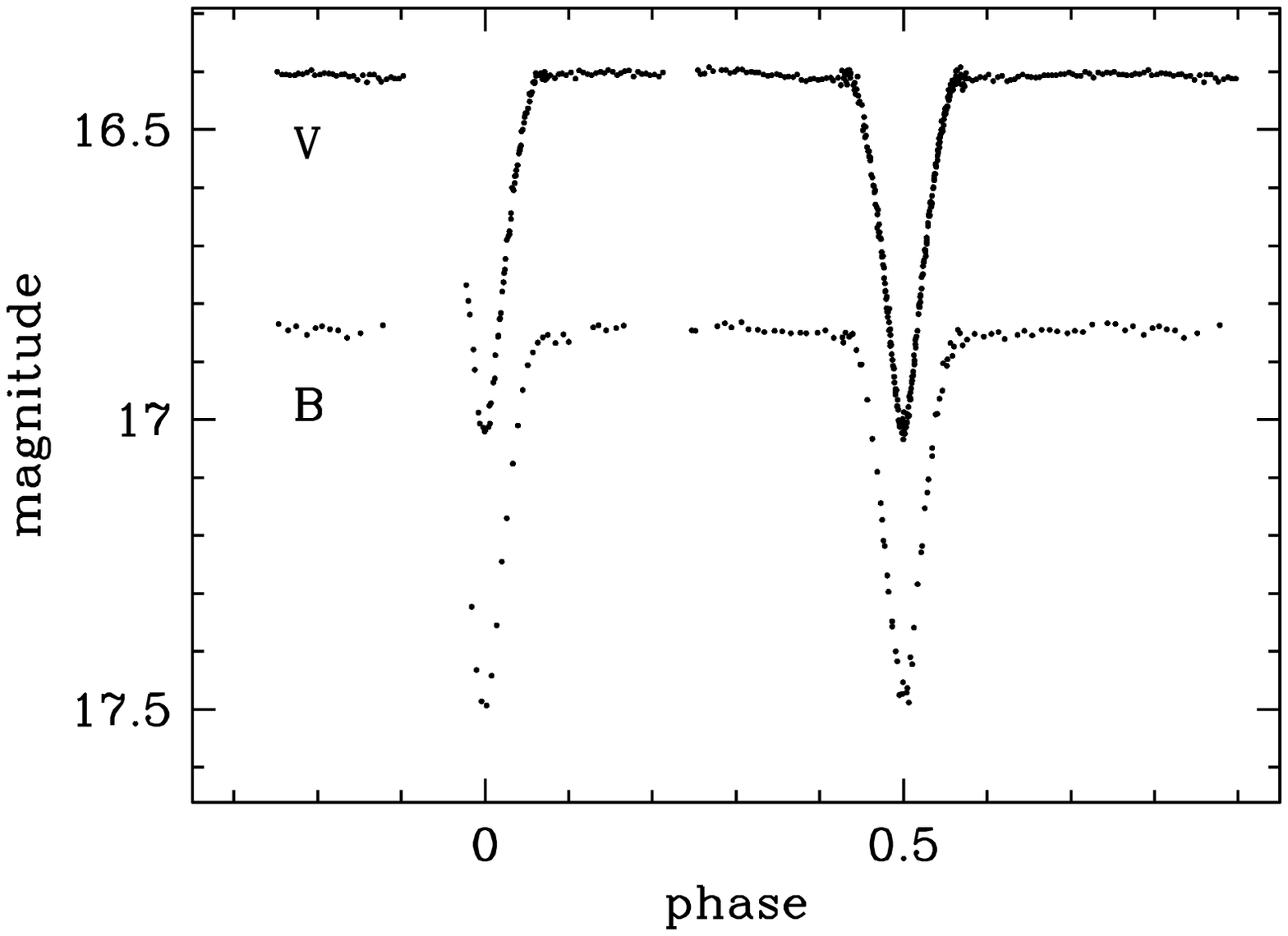}}
\FigCap{Phased $BV$ light curves of NV~CMa.}
\end{figure}
%
%fig 4
\begin{figure}[htb]
\centerline{\includegraphics[height=80mm,width=120mm]{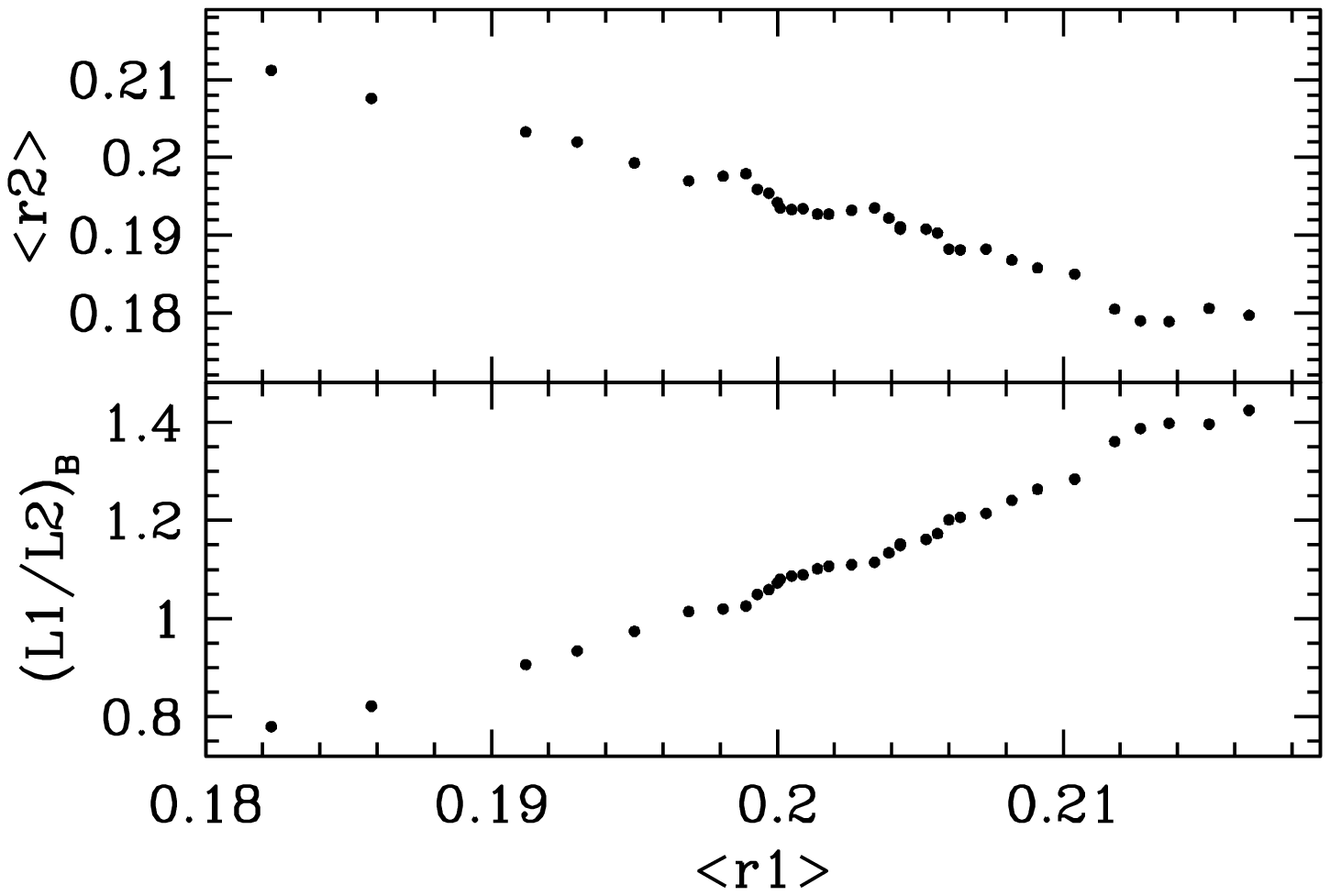}}
\FigCap{Dependence of the luminosity ratio $(L_{1B}/L_{2B})$ and
relative radius $<r_{2}>$ on the assumed radius $<r_{1}>$.}
\end{figure}
%
%fig 5
\begin{figure}[htb]
\centerline{\includegraphics[height=50mm,width=120mm]{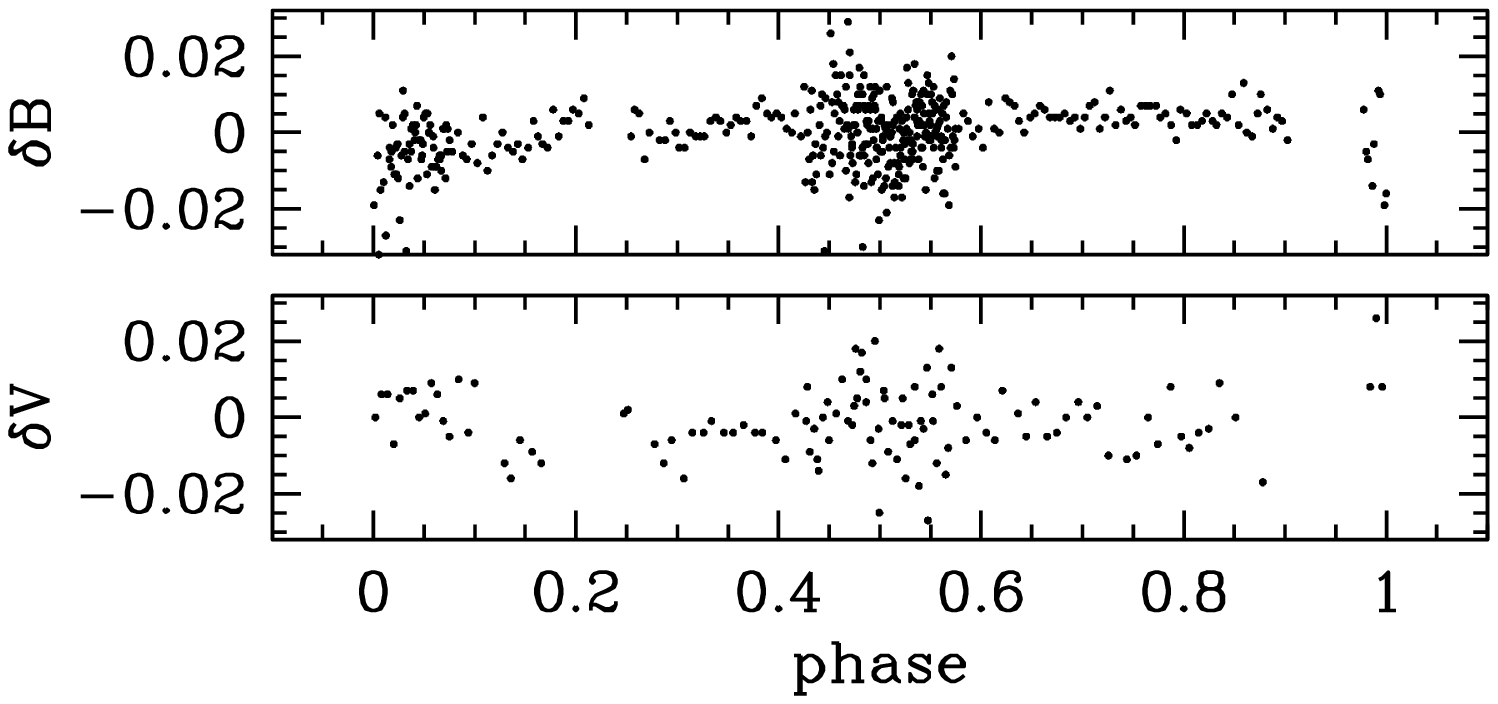}}
\FigCap{The residuals for the fit corresponding to the light curve solution
listed in Table~5.}
\end{figure}
%
%fig 6
\begin{figure}[htb]
\centerline{\includegraphics[height=100mm,width=120mm]{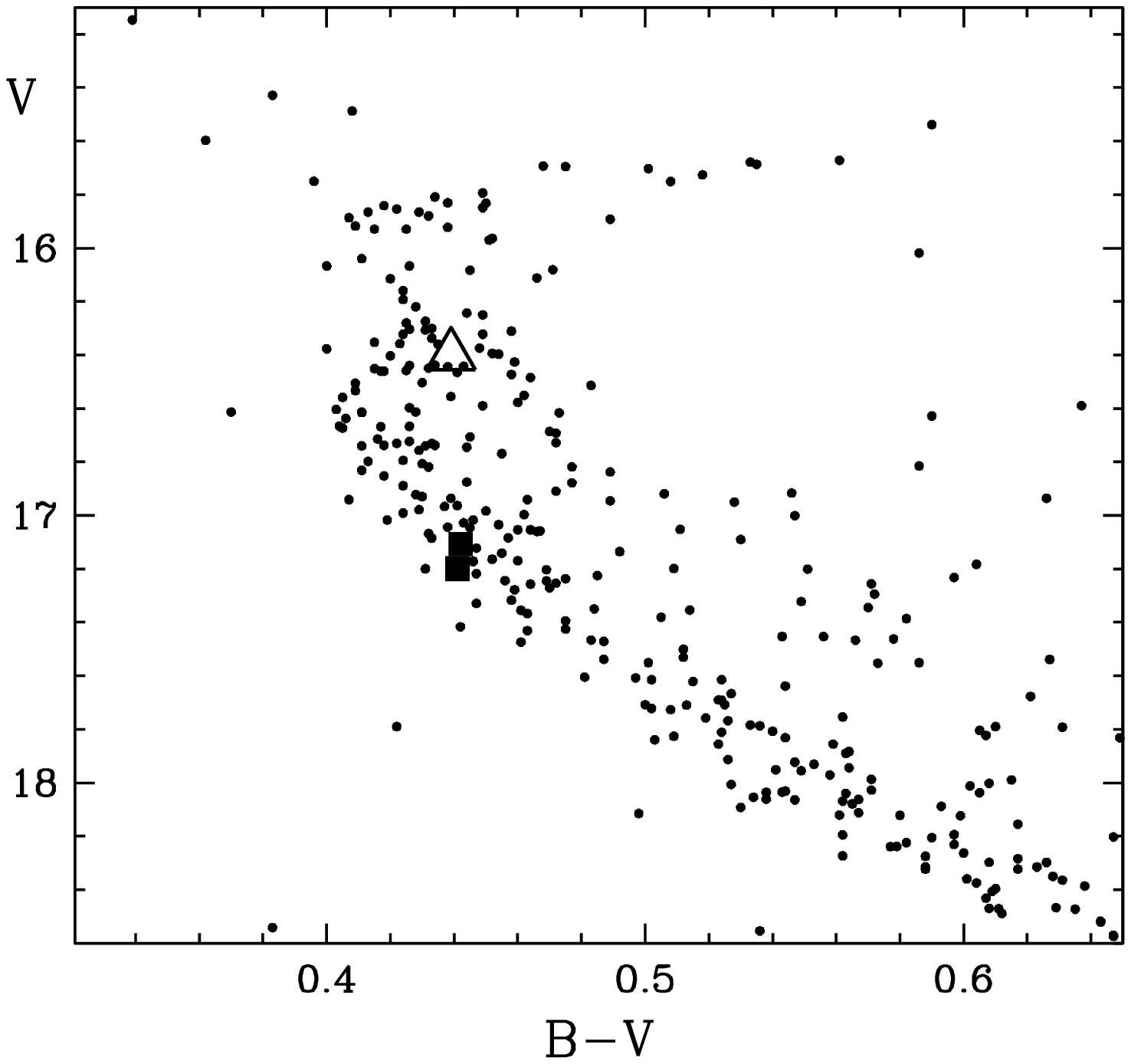}}
\FigCap{Color-magnitude diagram for the turnoff region of NGC~2243. The location
of NV~CMa is marked with an open triangle. The squares denote the locations of
individual components of the binary.}
\end{figure}
%
%fig 7
\begin{figure}[htb]
\centerline{\includegraphics[height=90mm,width=120mm]{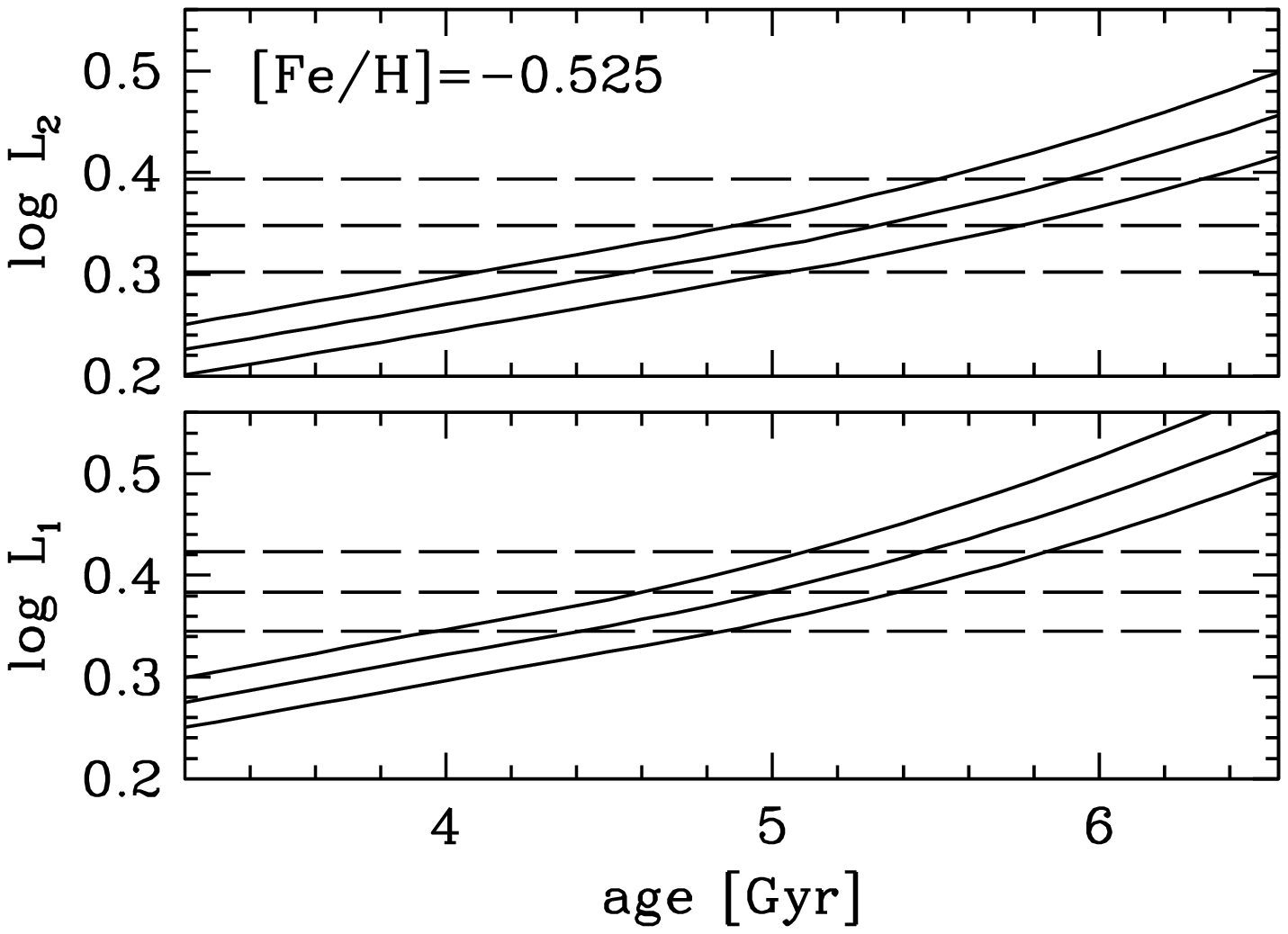}}
\FigCap{Theoretical age-luminosity relations for masses
$M_{1}=1.089\pm 0.010 M_{\odot}$ (lower) and
$M_{2}=1.069\pm 0.010 M_{\odot}$ (upper). Horizontal dashed lines mark $1\sigma$
ranges for the observed luminosities of the components of NV~CMa. The solid lines represent the mass $\pm~1~\sigma$.}
\end{figure}
%
%fig 8
\begin{figure}[htb]
\centerline{\includegraphics[height=90mm,width=120mm]{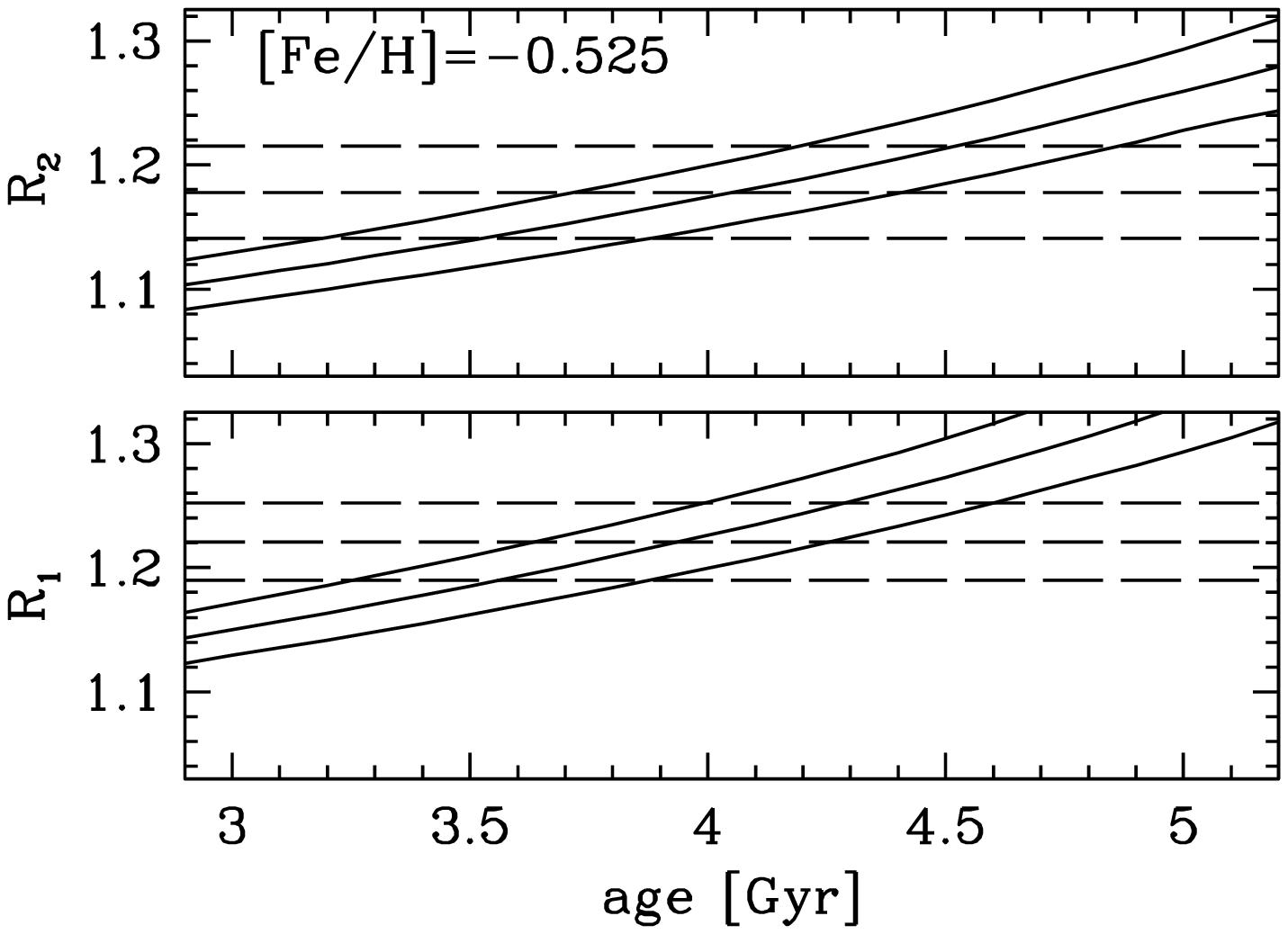}}
\FigCap{Theoretical age-radius relations for masses
$M_{1}=1.089\pm 0.010 M_{\odot}$ (lower) and
$M_{2}=1.069\pm 0.010 M_{\odot}$ (upper). Horizontal dashed lines mark $1\sigma$
ranges for the observed radii of the components of NV~CMa. The solid lines represent the mass $\pm~1~\sigma$}
\end{figure}
%
%fig 9
\begin{figure}[htb]
\centerline{\includegraphics[height=80mm,width=120mm]{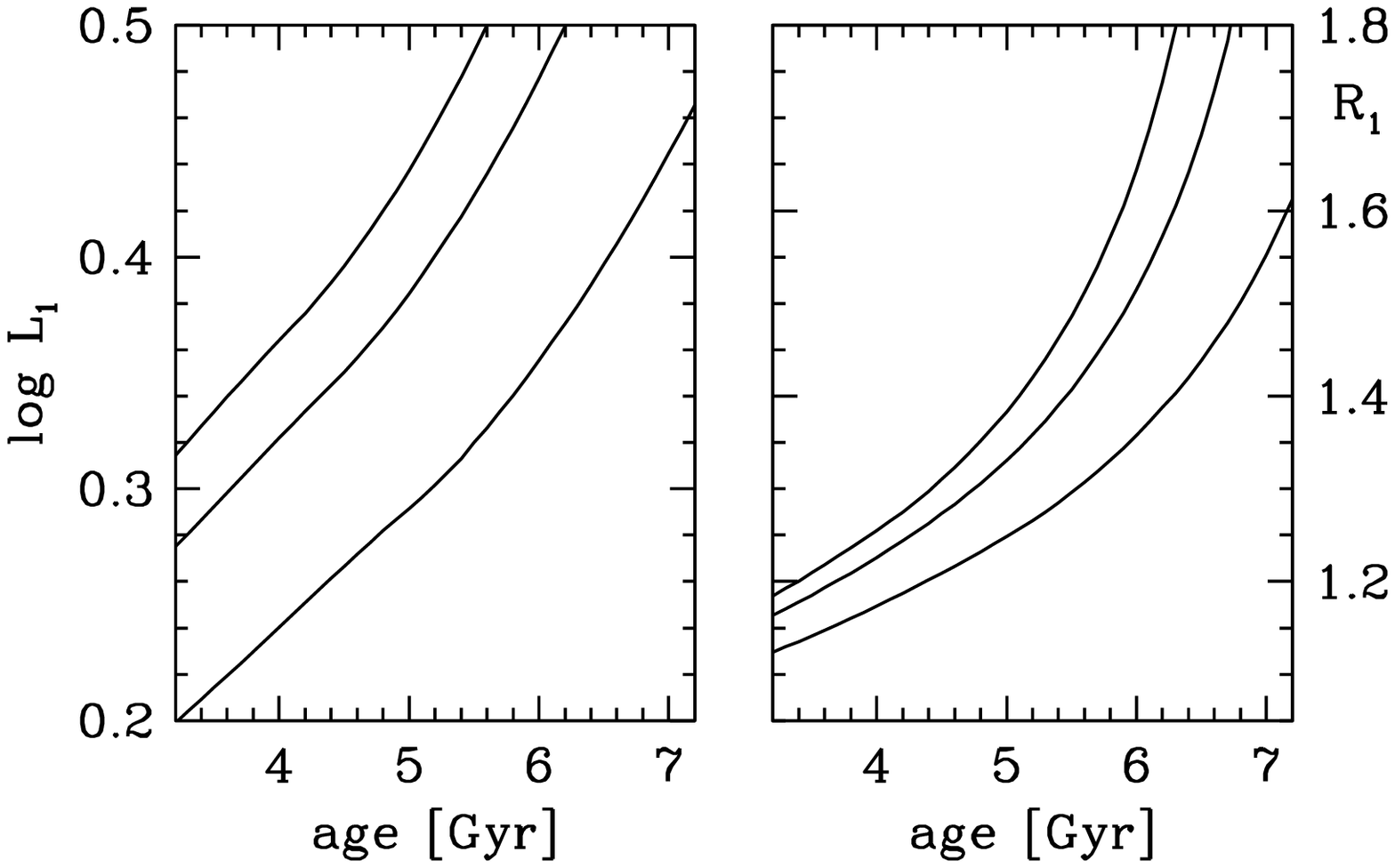}}
\FigCap{Theoretical age-luminosity and age-radius relations for
a star with mass $M=1.089 M_{\odot}$ and for metallicity
${\rm [Fe/H]=-0.606}$ (upper curve), ${\rm [Fe/H]=-0.525}$ (middle curve)
and ${\rm [Fe/H]=-0.397}$ (lower curve).}
\end{figure}
\Acknow{ JK and WP were supported by the grant
1~P03D~001~28 from the Ministry
of Science and Information Society Technologies, Poland. Research of
SMR is supported by a grant from the Natural Sciences and Engineering 
Council of Canada while IBT is 
supported by NSF grant AST-0507325. We thank Alexis Brandeker who
obtained one of the spectra of NV~CMa used here. It is also a pleasure to
thank Tomek Plewa for enlightening hints on MINGA usage.
}
\end{document}